\documentclass[reprint,%
 amsmath,amssymb,
 aps, prl
]{revtex4-1}
\usepackage{braket}
\usepackage{graphicx}
\usepackage{dcolumn}
\usepackage{bm}

\newcommand{\pot}{${}^{39}\mathrm{K}~$}

\begin{document}


\title{Observation of the algebraic localization-delocalization transition in a 1D disordered potential with a bias force }

\author{G. Berthet, L. Lavoine, M.K. Parit, A. Brolis, A. Boiss\'e, T. Bourdel}
\affiliation{Laboratoire Charles Fabry, Institut d'Optique Graduate School, CNRS, Universit\'e Paris-Saclay, 2 av. A. Fresnel, 91127 Palaiseau Cedex, France}

\date{\today}

\begin{abstract}

In a one-dimensional (1D) disordered potential, quantum interferences leading to Anderson localization are ubiquitous, such that all wave-functions are exponentially localized. Moreover, no phase transition toward delocalization is expected in 1D. This behavior is strongly modified in the presence of a bias force. We experimentally study this case, launching a non-interacting $^{39}$K Bose-Einstein condensate in a 1D disordered potential induced by a far-off-resonance laser speckle, while controlling a bias force. In agreement with theoretical predictions, we observe a transition between algebraic localization and delocalization as a function of our control parameter that is the relative strength of the disorder against the bias force. We also demonstrate that the initial velocity of the wave-packet only plays a role through an effective disorder strength due to the correlation of the disorder.

\end{abstract}

\maketitle



Adding a bias force is a quite natural way to probe the transport properties of quantum systems, a subject of broad interest that can be in particular addressed with atomic quantum gases thanks to their high degree of control and versatility \cite{Chien15}. For example, Bloch oscillations has been measured through the addition of a constant force to atoms in periodic potential induced by an optical lattice \cite{BenDahan96}.  A force applied to a harmonic trap is equivalent to a trap displacement. The response to such a displacement  permits to reveal the fluid or insulating behavior of atomic systems. In 1D interacting Bose gases, the pinning transition by an optical lattice \cite{Haller10} or the insulating transition in quasi-disordered optical lattice \cite{Tanzi13, DErrico14} have been studied in this manner. More recently, transport in quantum gases is also studied in junction-type setup more analogous to condensed-matter systems: two reservoirs with different chemical potentials are connected through a constriction. For example, in a gas of fermions, the quantization of conductance through a quantum point contact \cite{Krinner15} and the superfluid to normal transition in a disordered thin film have been observed \cite{Krinner13}.

In our work, we focus on the transport of non-interacting particles in disordered media. Without a bias  force, quantum interferences between multiple paths lead to Anderson localization \cite{Anderson58} whose signature is an exponential decay in space of single particle wave-function \cite{Sanchez-Palencia07}. This phenomenon is ubiquitous in wave/quantum physics and it has been observed in many physical contexts \cite{Lagendijk09} including condensed-matter \cite{Gomez05} and ultra-cold atoms \cite{Billy08, Roati08, Chabe08}.  One-dimensional truly disordered systems are always localized \cite{Abrahams79}, contrary to the 3D case where a phase transition between localized and extended single particle wave-functions takes place as a function of the disorder strength \cite{Lopez12, Jendrzejewski12, Kondov11}.  

The localization properties of 1D disordered systems are modified in the presence of a bias force. Theoretical studies predict a transition from algebraic localization to delocalization as a function of a single control non-dimensional parameter $\alpha$ which is the ratio of the force to the disorder strength \cite{Crosnier17, Crosnier18}. Physically, $\alpha$ is the relative energy gain $\Delta E/E$ of a particle of energy $E$  when moving over a localization length. Interestingly, in a 1D white noise disorder, this quantity is independent of $E$ as the localization length is proportional to $E$. If $\alpha$ is small, the force does not considerably change the localization behavior of the particle while for large $\alpha$ its dynamics is severely affected leading to delocalization. 

This localization-delocalization transition is predicted in the infinite time limit for white noise disorder \cite{Crosnier18}. 
In a correlated disorder, as the one produced  from a far-off-resonance laser speckle \cite{Clement06}, the situation is more complicated. Speckles have no Fourier component beyond a spatial frequency $2k_c$. As a consequence, back-scattering and localization are not expected in the framework of Born approximation for atoms with wavevectors $k > k_c$ \cite{Billy08, Lugan09}. Since localized wave-functions always have a small fraction at long distance corresponding to large energies and momenta in the presence of a bias force, we thus expect correlation-induced delocalization at infinite time. However, signatures of the algebraic localization-delocalization transition are predicted to be observable at transient times \cite{Crosnier18}. 

In this paper, we report on the observation of the algebraic localization-delocalization transition with cold-atoms propagating in a one dimensional disordered potential in the presence of a controlled bias force. We experimentally show that the non-dimensional parameter $\alpha$ is the only relevant parameter to describe the transition. We notice that the initial velocity of the quantum wave packet only plays a role through the correlation of the disordered potential, showing that the transition is intrinsically energy independent. In the localized regime, we demonstrate an algebraic decay of the density and measure the corresponding decay exponent as a function of $\alpha$. At large disorder strength, a saturation of the exponent is explained by an effect of the disorder correlation. 

\begin{figure}[htb!]
\centering
\includegraphics[scale=0.93]{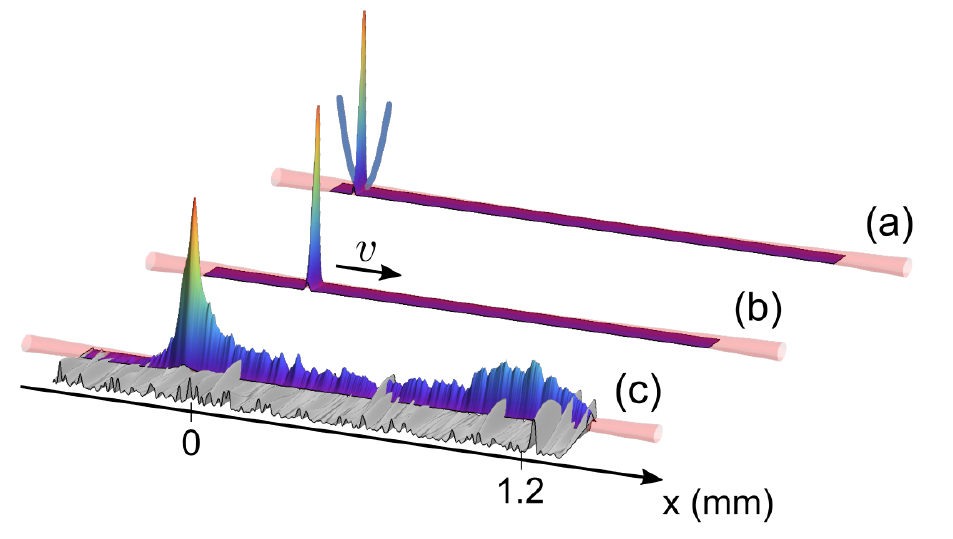}
\caption{(Color online) Schematic of the experimental sequence. A non-interacting atomic Bose-Einstein condensate is produced in a crossed trap (a). The cloud is then launched into a 1D tube with a controlled force (b). The cloud can be accelerated to a velocity $v$. A constant force is then applied for a time $\tau$ to the system while a 1D speckle at $532~\mathrm{nm}$ is shone on the atoms (c). An average density profile over 8 disorder realizations is shown.}
\label{fig:SchemaManip}
\end{figure}

We first produce a \pot condensate. In contrast to our previous works \cite{Salomon14, Lepoutre16, Boisse17},  after optical cooling, we pump the atoms to the $\ket{F=2,m_F=2}$ state and load a magnetic trap which serves as a reservoir for loading a tightly confining optical trap \cite{Landini12}. The evaporation is then pursed in the $\ket{F=1,m_F=1}$ state in the vicinity of the $403.4(7)~\mathrm{G}$   Feshbach resonance \cite{DErrico07} until condensation is reached. The magnetic field is then ramped down close to the scattering length zero crossing at $350.4(4)$\,G in order to be in a non-interacting regime. The scattering length is then $-0.2 \pm 0.2\,a_0$ with $a_0$ the Bohr radius. The final trap is made of two horizontal far-detuned laser beam at 1064\,nm and 1550\,nm. Its oscillation frequencies are $18 \times 124 \times 124\,$Hz. When the 1550\,nm beam is turned off, the longitudinal confinement (along $\hat{x}$) is removed and the atoms are free to evolve in the 1064\,nm trap whose radial frequencies are $\omega_\perp/2\pi=124\,$Hz \cite{footnoteBloch} (see fig.\,\ref{fig:SchemaManip}). The residual trapping frequency in the longitudinal direction is precisely canceled to 0.0(5)\,Hz by using the magnetic field curvature induced by two pairs of magnetic coils. We also control a longitudinal force, characterized by an acceleration $a$, through a magnetic field gradient produced by an off-centered additional coil. 

A 1D disorder potential is shone on the atoms for a time $t\sim \tau$ while they propagate in the one-dimensional trap under the action of the constant force. The atoms are then observed by resonant fluorescence imaging after a 1\,ms time of flight corresponding to the time needed to properly turn off the magnetic field. For intermediate disorder strength, the density profile (see fig.\,\ref{fig:SchemaManip}c) can be separated into three components. Some atoms do not move substantially. The fraction of localized atoms is directly retrieved by integration over a zone of $\sim$300\,$\mu$m around the initial position. Some atoms are scattered by the disorder and are accelerated at a later time. Finally, some atoms behave ballistically. The propagation time is adjusted between 90\,ms and  500\,ms, such in the absence of disorder the ballistic atoms have travelled about 1\,mm and are still in the camera field of view. In this case, the velocity spread of the atomic cloud is measured to be $\Delta v=0.37(6)\,$mm.s$^{-1}$. 

The disorder is produced from a laser speckle at 532\,nm, its effect on the atoms is the one of a conservative potential. The 532\,nm laser beam propagates along the $\hat{z}$ axis and passes through a diffusing plate. The amplitude $V_R$ of the speckle potential is directly proportional to the laser intensity and corresponds to both the mean value and the standard deviation of the exponential probability distribution of the potential \cite{Clement06}. On the diffusing plate, the intensity distribution is elliptical with the major axis along $\hat{x}$ and the minor axis along $\hat{y}$. This produces an anisotropic disorder. The autocorrelation widths are $4.7\,\mu\mathrm{m}$ along $\hat{y}$ and  $4.3\,\mu\mathrm{m}$ along $\hat{z}$ (half-width at $1/\sqrt{e}$). Both values largely exceeds the ground state extension of the harmonic oscillator $\sqrt{\hbar/2 m \omega_\perp}=1.0\,\mu\mathrm{m}$, where $m$ is the atomic mass and $\hbar$ the reduced Planck constant. The disordered potential can thus be considered as one-dimensional for the atoms propagating along $\hat{x}$. In this direction, the power spectral density is precisely measured. It takes zero values for wavevectors larger than $k_c=2/\sigma$ with $\sigma=0.34\,\mu$m. For the comparison with a white noise potential, the power spectral density at zero momentum that can be written $\tilde{C}(k=0)=c V_R^2 \pi \sigma$, where $c=1.26$ is a numerical constant that we have experimentally measured on the speckle pattern. It would be 1 for a uniform intensity on the diffusing plate. 

Let us now discuss, the relevant energy scales of our experiment. The correlation length of the disorder gives the correlation energy $E_\sigma=\hbar^2/2 m \sigma^2=(2 \pi \hbar)\times1.1$\,kHz. In order for the correlation of the disorder to play little role, it has to be the largest energy scale. It has to be compared to the disorder amplitude $V_R$ and to the energy scale $E_a=\hbar^{2/3} m^{1/3} a^{2/3}$ associated with the acceleration $a$. Experimentally, $E_a/(2\pi\hbar)$ takes values between 7 and 33 Hz and $V_R/(2\pi\hbar)$ is varied between zero and 200\,Hz. The relevant energy scale for the disorder when comparing with a white noise potential is rather $V^*=\hbar^{-2/3} m^{1/3}  \tilde{C}(0)^{2/3}$, which is lower than $V_R$. The non-dimensional parameter $\alpha$ is then defined as $\alpha=(E_a/V^*)^{3/2}=\hbar^2 a/ \tilde{C}(0)$ such that the localization-delocalization transition is expected for $\alpha=1$ \cite{Crosnier18}. Another energy scale $E_t=\hbar/\tau$ can be associated to the propagation time. Its value is chosen to be below the value of $E_a$ by a factor of the order of 15 to 20, such that propagation, scattering and localization phenomena have enough time to set in. 


\begin{figure}[htb!]
\centering
\includegraphics[scale=0.35]{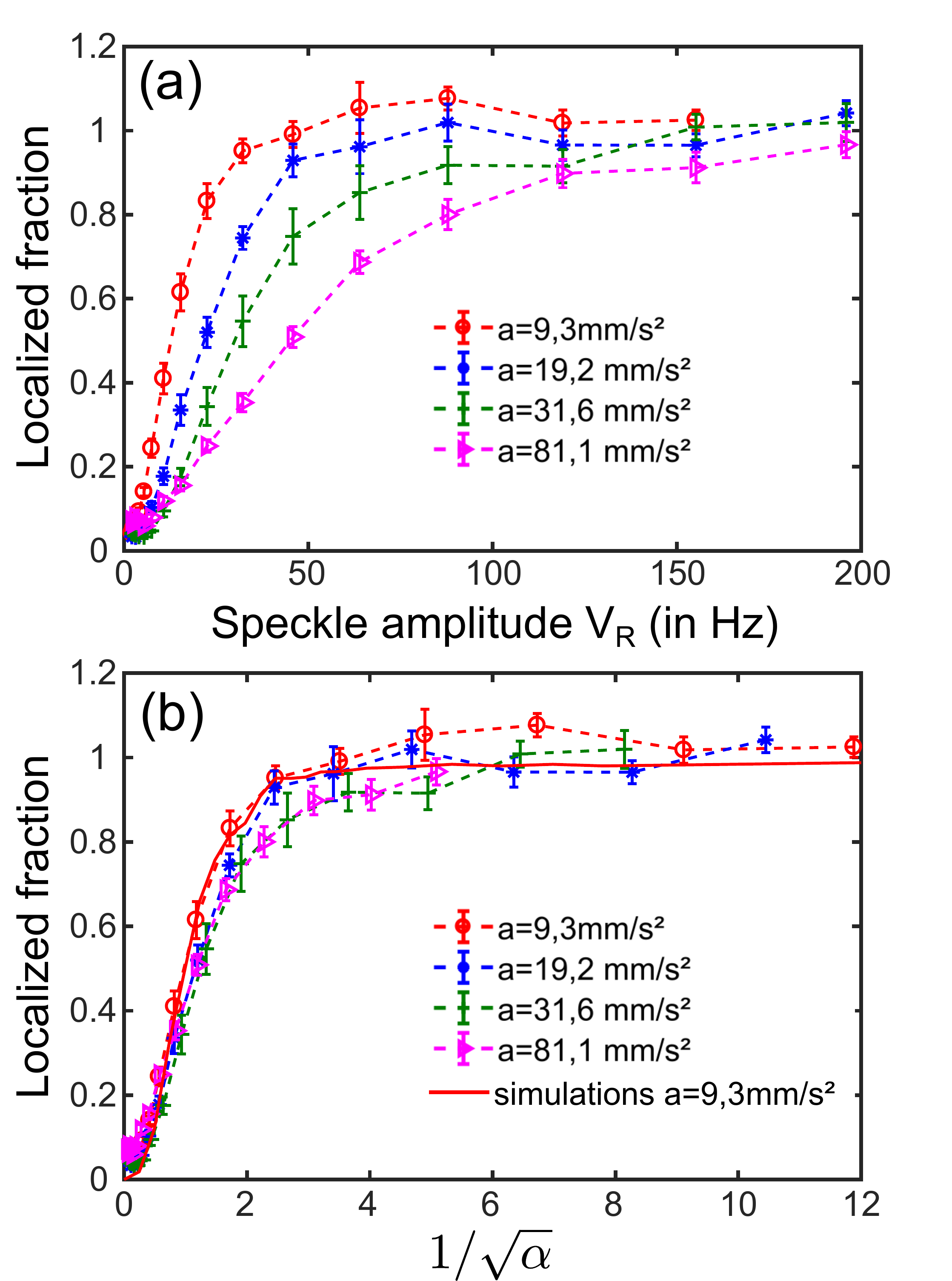}
\caption{(Color online) Localized atomic fraction as a function of the speckle amplitude $V_R$ (a) and $1/\sqrt{\alpha}$ (b) obtained for $4$ different values of the applied acceleration. Each curve is averaged over $8$ realizations of the disorder potential. The error bars are  the standard deviations of the measured localized fractions. Experimental values above 1 are due to small errors in background substraction. The continuous red line is the average of numerical simulations of the 1D Schr\"odinger equation.}
\label{fig:resultLoc}
\end{figure}

We first study the case of atoms entering the disorder without initial velocity. The measurements of the localized fraction are performed for four different values of the acceleration $a=9.30~\mathrm{mm.s^{-1}}$, $a=19.2~\mathrm{mm.s^{-1}}$, $a=31.6~\mathrm{mm.s^{-1}}$ and $a=81.1~\mathrm{mm.s^{-1}}$ respectively associated with propagation times $\tau=460\,$ms, $\tau=320\,$ms, $\tau=280\,$ms, $\tau=90\,$ms. For each acceleration, the speckle amplitude $V_R$ is scanned between zero and $V_R/(2\pi \hbar)=200\,$Hz. The sequence is then repeated over $8$ different realizations of the speckle potential in order to average our results. Each realization is obtained by moving the diffusing plate. The measured localized fractions are reported in fig. \ref{fig:resultLoc}. For small values of the speckle amplitude $V_R$ (see fig. \ref{fig:resultLoc}(a)) the system is mostly delocalized whatever the acceleration. On the contrary, all the atoms are localized when the speckle strength is high. One notes that the lower the acceleration the faster the localized fraction increases with $V_R$. 

Rescaling the horizontal axis with the dimensionless parameter $1/\sqrt{\alpha}$, which is proportional to $V_R$ leads to a clear collapse of the data within the error bars (see fig. \ref{fig:resultLoc}(b)). This indicates that $\alpha$, which compares the acceleration to the disorder strength, is the only relevant parameter. If the localized-delocalized transition point is defined for a localized fraction equal to $0.5$, this corresponds to $\alpha=1.0(3)$ (fig. \ref{fig:resultLoc}(b)). We can compare these results with disorder average numerical simulations of the 1D Schr\"odinger equation for our parameters. We use these simulations to calibrate $V_R$ as a function of the optical power with a $15\%$ uncertainty.

\begin{figure}[htb!]
\centering
\includegraphics[scale=0.35]{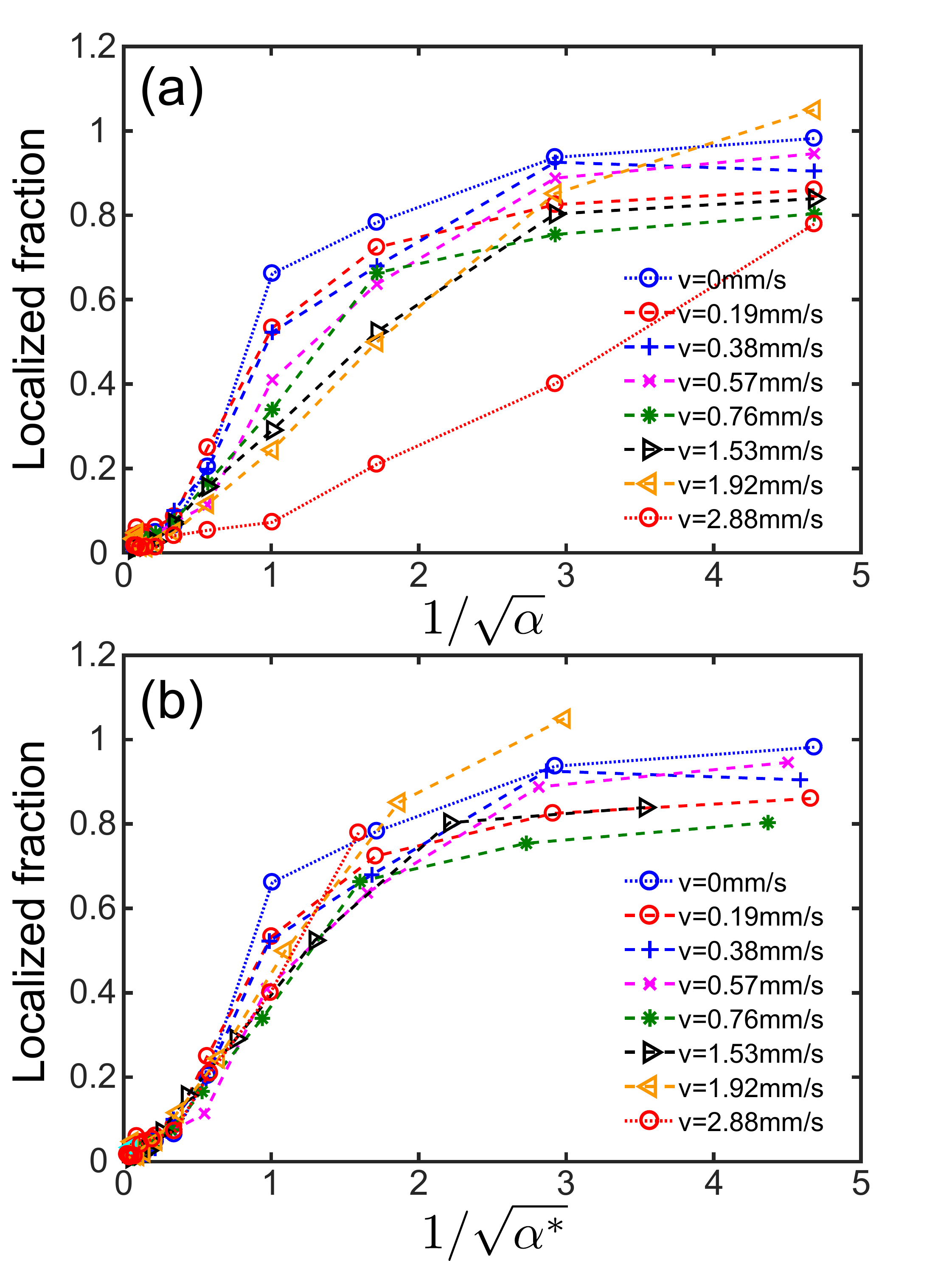}
\caption{(Color online) Localized fractions of the non-interacting cloud of atoms as a function of $1/\sqrt{\alpha}$ (a) and $1/\sqrt{\alpha^*}$ (b) for different values of the initial velocity.}
\label{fig:resultLocAvecVit}
\end{figure}

We now consider the case of atoms entering the disorder with an initial positive velocity. The initial velocity $v$ is first set by applying an acceleration to the atoms during $10~\mathrm{ms}$ (see fig.\ref{fig:SchemaManip}). The acceleration is then changed to its final value equal to $a=19.2~\mathrm{mm.s^{-1}}$ and the speckle is simultaneously shone on the atoms. The diffusion time in the disordered potential is set to $280~\mathrm{ms}$. The measured localized fractions are presented in fig.\,\ref{fig:resultLocAvecVit}  \cite{Boisse17}. For the highest velocity ($v=\hbar k/m=2.88~\mathrm{mm.s^{-1}}$), $k \sigma_x \simeq 0.6$ and the value of the disorder potential's power spectrum $\tilde{C}(2k)$ is significantly reduced as compared to  $\tilde{C}(0)$. This correlation effect is responsible for the non-collapsing behavior when the localized fractions are plotted as a function of $1/\sqrt{\alpha}$ (see fig.\ref{fig:resultLocAvecVit}(a)). Plotting as a function of $1/\sqrt{\alpha^*}$, where $\alpha^*=\hbar^2 a/\hat{C}(2k)$ takes into account an effective disorder strength at the atomic momentum $k$, leads to a much better collapse of the data on the curve at zero velocity for which $\alpha^*=\alpha$ (see fig.\ref{fig:resultLocAvecVit}(b)). This indicates that $\alpha^*$ is the only relevant parameter and that the initial velocity only plays a role through the correlation of the disorder \citep{Crosnier18}. The transition point corresponding to a localized fraction of $0.5$ is obtained for $\alpha^*=1.0(4)$ (fig. \ref{fig:resultLocAvecVit}(b)).

\begin{figure}[htb!]
\centering
\includegraphics[scale=0.35]{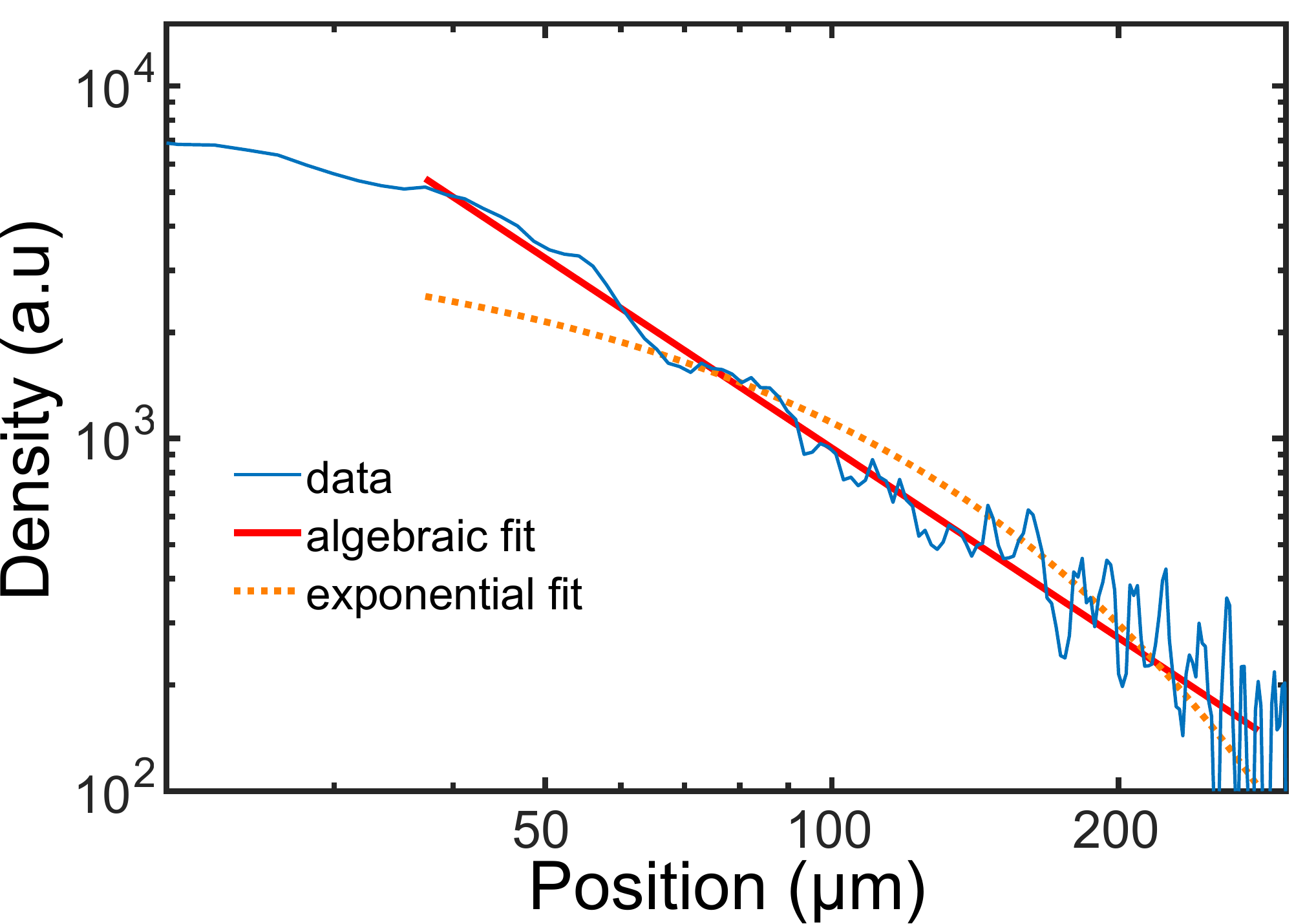}
\caption{(Color online) Averaged localized density profile as a function of the atom displacement in log-log scale for $a=9.2\,$mm.s$^{-2}$, $\alpha=0.03$ and no initial velocity. Log-log scale permits to show the algebraic dependance as a straight line. The algebraic (red continuous line) and exponential (orange dotted line) fits are displayed. The fit with the power law $1/x^\beta$ leads to $\beta=1.79 \pm 0.10$, where the error bar is evaluated from reduced data sets of 4 speckle realizations.}
\label{fig:exempleExposant3}
\end{figure}

Finally, we perform a more careful analysis of the localized profiles for the specific case of $a=19.2\,$mm.s$^{-2}$. We find that singles profiles are quite noisy with modulations that are stable for a given speckle realization. This shows that the localization profile is not a self averaging quantity. We thus average results for 8 different realizations of the speckle in order to obtain smooth localization profiles (Fig.\,\ref{fig:exempleExposant3}). An algebraic scaling is visible for distances between 40\,$\mu$m and 300\,$\mu$m (a straight line in log-log scale), whereas an exponential decay in this region does not fit the data \citep{Crosnier18}. Below 40\,$\mu$m, there is an effect of the initial size of the cloud. Above 300\,$\mu$m, the signal to noise ratio is low. 

\begin{figure}[htb!]
\centering
\includegraphics[scale=0.35]{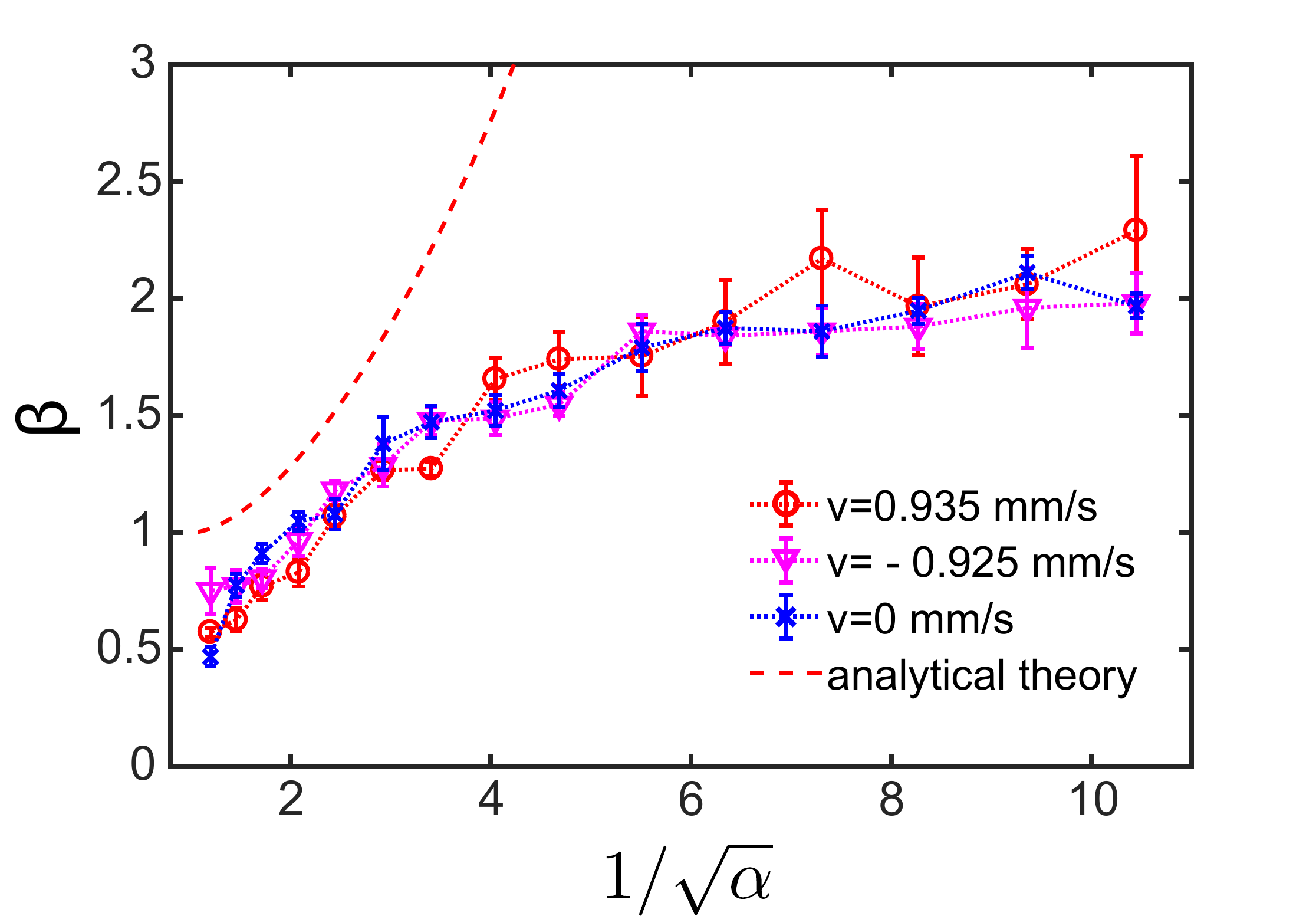}
\caption{(Color online) Values of $\beta$, obtained by fitting to the localized density profile with a power law, as a function of $1/\sqrt{\alpha}$ for $3$ different values of the initial velocity.}
\label{fig:exposant}
\end{figure}

We now study the algebraic decay coefficient $\beta$ as a function of the disorder strength or equivalently as a function of $\alpha$ (see Fig.\,\ref{fig:exposant}). The coefficient is found to increase with increasing speckle as can be expected for stronger localization effects. Moreover, we show that the behavior is not modified by a small positive or negative initial velocity ($v=\pm0.935~\mathrm{mm.s^{-1}}$ and $\alpha\simeq\alpha^*$ in this case) as expected theoretically. However, our results does not match with the infinite time white noise analytical exponents from \cite{Crosnier18} (dashed line in Fig.\,\ref{fig:exposant}). The experimental values of $\beta$ are always below the theoretical one and the analytical curve does not show the observed saturation $\beta$ at large speckle amplitudes. We interpret both effects as due to the correlation of our speckle potential. At a distance of 300\,$\mu$m, $k \sigma \sim 0.7$ and the disorder correlation together with the finite evolution time is likely to affect the measured $\beta$ exponent. The saturation can be interpreted as a combined strong disorder and correlation effect. A strong disorder ($V_R \sim E_\sigma$) broadens the initial wavevector distribution such that $k$-vectors around $k_c$ are populated due to the sudden switching-on of the disorder. This can be understood from the broad width of the spectral functions \cite{Prat16, Volchkov18}. In this case, even in the absence of a force, the population close to $k_c$ is responsible for an algebraic decay of the localized density profiles with a coefficient 2 \cite{Billy08} in agreement with our measurement. In this regime, the acceleration plays no role as the dominant energy scales are $V_R$ and $E_\sigma$.


In conclusion, we report on the observation of the algebraic localization-delocalization transition with ultra-cold matter waves in the presence of a controlled bias force. The localized fraction of atoms only depends on a non-dimensional parameter which is the ratio of the force to the disorder strength. The initial velocity only plays a role through a rescaling of the disorder strength due the correlation of the disordered potential and the localization-delocalization appears as an energy independent phenomenon. Algebraic localization is observed. For the algebraic decay exponents, a large discrepancy with the analytical white noise theory is interpreted as a consequence of a correlated and strong disorder. 

Whereas adding a bias force is a natural tool to study transport in both condensed-matter and ultra-cold atomic disorder systems, our result show that it can have important consequences \cite{Crosnier17}. More specifically, our results pave the way to the study 1D $interacting$ bosons in the presence of disorder, when modifying the scattering length. For example, a finite temperature localization-delocalization phase transition has been predicted due to many-body localization effects \cite{Aleiner10}. The study of the phase diagram of 1D strongly interacting disordered Bose systems is also of interest \cite{Giamarchi88, DErrico14}.

\begin{acknowledgments}
This research has been supported by CNRS, Minist\`ere de l'Enseignement Sup\'erieur et de la Recherche, Labex PALM, ERC senior grant Quantatop, Region Ile-de-France in the framework of DIM Nano-K (IFRAF) and DIM Sirteq, EU - H2020 research and innovation program (Grant No. 641122 - QUIC), Paris-Saclay in the framework of IQUPS, Simons foundation (award number 563916: localization of waves). 
\end{acknowledgments}


\begin{thebibliography}{0}
\bibitem{footnoteBloch} 
 Due to its normal incidence to the science chamber and its large coherence length, the $1064~\mathrm{nm}$ laser beam produces a sub-nanoKelvin residual periodic lattice on the atoms that can lead to Bloch oscillations \cite{BenDahan96}. We thus limit the propagation time of our experiment in order that the atoms never reach the necessary velocity of $9.6~\mathrm{mm.s^{-1}}$ corresponding to half of the lattice wavevector.
\end{thebibliography}
\end{document}